\documentclass[twocolumn]{aastex631}

\usepackage{amsmath}
\usepackage{tabularx}
\usepackage{mathrsfs}
\usepackage{xcolor}
\usepackage{CJK}
\usepackage[caption=false]{subfig}

\shorttitle{Binary planet dynamics within clusters}
\shortauthors{Huang et al.}

\graphicspath{{./}{figures/}}

\begin{document}

\begin{CJK*}{UTF8}{gbsn}
\title{Dynamics of Binary Planets within Star Clusters}

\author[0000-0003-1215-4130]{Yukun Huang (黄宇坤)}
\affiliation{National Astronomical Observatory of Japan, 2-21-1 Osawa, Mitaka, Tokyo 181-8588, Japan}
\affiliation{Department of Astronomy, Tsinghua University, Beijing 100084, China}
\author[0000-0003-4027-4711]{Wei Zhu (祝伟)}
\affiliation{Department of Astronomy, Tsinghua University, Beijing 100084, China}
\author[0000-0002-5486-7828]{Eiichiro Kokubo (小久保英一郎)}
\affiliation{National Astronomical Observatory of Japan, 2-21-1 Osawa, Mitaka, Tokyo 181-8588, Japan}

\begin{abstract}

We develop analytical tools and perform three-body simulations to investigate the orbital evolution and dynamical stability of binary planets within star clusters. Our analytical results show that the orbital stability of a planetary-mass binary against passing stars is mainly related to its orbital period. Critical flybys, defined as stellar encounters with energy kicks comparable to the binary binding energy, can efficiently produce a wide range of semimajor axes ($a$) and eccentricities ($e$) from a dominant population of primordially tight JuMBOs. The critical flyby criterion we derived offers an improvement over the commonly used tidal radius criterion, particularly in high-speed stellar encounters. Applying our results to the recently discovered Jupiter-Mass Binary Objects (JuMBOs) by the James Webb Space Telescope (JWST), our simulations suggest that to match the observed $\sim$9\% wide binary fraction, an initial semimajor axis of $a_0 \sim 10$--20~au and a density-weighted residence time of $\chi \gtrsim 10^4$~Myr~pc$^{-3}$ are favored. These results imply that the JWST JuMBOs probably formed as tight binaries near the cluster core.

\end{abstract}


\section{Introduction} \label{sec:intro}
\end{CJK*}

With an estimated age of $<$1~Myr and a distance at $\approx$390~pc \citep{Apellaniz.2022}, the Trapezium cluster is one of the youngest star clusters near the Sun. It is located at the center of the Orion Nebular Cluster and has one of the densest stellar environments \citep{McCaughrean.1994}. The extreme conditions of the Trapezium Cluster have made it an ideal target for studying the formation of stars and planetary-mass objects (PMOs), both theoretically and observationally \citep[e.g.,][]{Prosser.1994, Muench.2008, Pearson.2023}.

The recent discovery by \citet{Pearson.2023} of 540 PMO candidates, including 40 Jupiter-Mass Binary Objects (JuMBOs), in the Trapezium cluster poses significant challenges to our conventional understanding of planet formation. In the core accretion model, giant planets form within circumstellar protoplanetary disks. The growth from planetesimals to a planetary core (several $M_\oplus$) can last up to a million years, followed by a gas accretion phase lasting up to $\sim$10~Myr \citep{Lissauer.1993, Pollack.1996}. This time span is already a factor of a few longer than the estimated age of the Trapezium Cluster. Furthermore, the timescale for ejecting planets out of embedded systems through planet--planet scatterings or stellar encounters is estimated to be another millions to hundreds of millions of years \citep{Juric.2008, Chatterjee.2008, Spurzem.2009, Nesvorny.2012}.

In addition, the abundance of these PMOs relative to stars is also surprisingly high. The Trapezium cluster is estimated to contain $\sim$2000 stellar members \citep{Morales-Calderon.2011}, so the relative abundance of PMOs is $>$$20\%$. This is above the limit on Jupiter-mass free-floating planets (FFPs) that were derived from microlensing surveys \citep{Mroz.2017}. It is also in tension with the substellar mass function derived from Euclid, albeit for a less dense environment \citep{Martin.2024}.

What is even more surprising is the discovery of 40 JuMBOs (with binary masses of 2--20~$M_{\rm J}$) at wide separations ($d=25$--400~au), which contribute $\sim$$9\%$ of all PMOs discovered in the Trapezium Cluster \citep{Pearson.2023}. Planet-mass binaries have previously been identified. For example, \citet{Best.2017} found 2MASS J11193254--1137466AB to be a pair of $\sim$4~$M_J$ objects with $d \approx 4$~au, and \citet{Beichman.2013} found WISE 1828+2650 to be a pair of 3--6~$M_J$ objects with $d < 0.5$~au. Oph 162225--240515 is a wide ($d\approx242$~au) binary that was originally thought to be in the planetary-mass regime \citep{Jayawardhana.2006}, but later studies have substantially revised the masses upward \citep{Luhman.2007}.

Despite these previous discoveries, the existence of such abundant ($\sim$$9\%$) low-mass binaries with wide separations has not been anticipated.

Explaining the formation of binary planets with wide separations presents a significant challenge. While two planets on closely aligned astrocentric orbits can theoretically be ejected as a binary pair during stellar encounters \citep{Wang.2024f}, recent numerical studies indicate that the production rate via this direct ejection mechanism is at least two orders of magnitude too low to account for the observed number and fraction of binary planets in the Trapezium Cluster \citep{Yu.2024j,Portegies-Zwart.2023}. Another potential formation pathway is the dynamical capture of free-floating planets in stellar clusters; however, this mechanism also yields formation rates that are at least two orders of magnitude lower than the observed fraction \citep{Perets.2012}. Tidal capture during repeated planet-planet scattering within planetary systems has also been proposed \citep{Ochiai.2014, Lazzoni.2023}. However, this process typically results in the formation of very tight binaries rather than wide ones.

Given the difficulties to form wide binary planets out of the protoplanetary disk (i.e., planet-like formation), star-like formation, in which binary planets form directly within the cluster, may be preferred \citep[see also][]{Portegies-Zwart.2023}, although the exact formation mechanism remains unclear and requires further investigation.

Once formed, the binary planets in the cluster may evolve as a result of dynamical interactions with passing stars. This is especially the case for such wide binaries as the JWST JuMBOs in the Trapezium Cluster. Due to the extremely high star number density ($n_\star\sim1\times10^4\ \text{pc}^{-3}$), the dynamical lifetime of a typical JWST JuMBO is comparable to the age of the cluster (see Section~\ref{sub:analytical}; see also \citealt{Portegies-Zwart.2023}). 
Therefore, it is probable that the wide binary planets seen by JWST were born with closer separations and then got softened by stellar flybys. These binary planets with initially close separations may then be the lower mass counterparts of binaries with L-, T-, and Y-type primaries, for which the binary fraction is 10--30\% and the binary separation peaks at 3--10~au \citep{Burgasser.2007, Joergens.2008}.

Regardless of the true nature of the recently discovered PMOs and JuMBOs by JWST, it is necessary to study the dynamical evolution of the planetary binaries in a dense cluster environment to explore their origins.

\section{Analytical Method}\label{sub:analytical}

The dynamical evolution of binary stars in star clusters has been extensively studied in the literature (e.g., \citealt{Heggie.1975}, \citealt{Heggie.2003}, and \citealt{Binney.2008}). One of the key conclusions from these seminal works is Heggie's law, which states that hard binaries become harder and soft binaries become softer over time due to multiple stellar encounters. However, the dynamics of binary planet--star interactions differ significantly from binary star--star interactions for the following reasons:

\begin{enumerate}
    \item An underlying assumption of \citet{Heggie.1975} is that the impact parameter must be smaller than the binary semimajor axis (i.e., $b < a$). This assumption is crucial because, for binaries with masses comparable to those of the passing stars, full three-body interactions are necessary to accurately describe their dynamics. In contrast, for binary planets with a total mass much smaller than that of the star, the star does not need to come as close to one component to perturb the binary. Instead, the primary perturbing force on a binary planet is the stellar tidal effect.
    \item The mass and binding energy of a binary planet are too small to meaningfully influence the star's motion. As a result, energy equipartition cannot be simply assumed to estimate the barycentric velocity of planetary bodies; otherwise, they would move $\sim$10$\times$ faster than stars. Numerical simulations of free-floating planets in star clusters, however, show a similar velocity distribution among planets and stars \citep{Wang.2015}.
\end{enumerate}

Given these fundamental differences, we aim to develop a straightforward analytical model focusing on the $m \ll M_\star$ interaction, where $m$ is the total mass of the binary and $M_\star$ is the much more massive third body. For simplicity, only equal-mass binaries are considered in this work.

\subsection{Effect of a Single Flyby}

\begin{figure}[htb!]
  \centering
  \includegraphics[width=0.8\columnwidth]{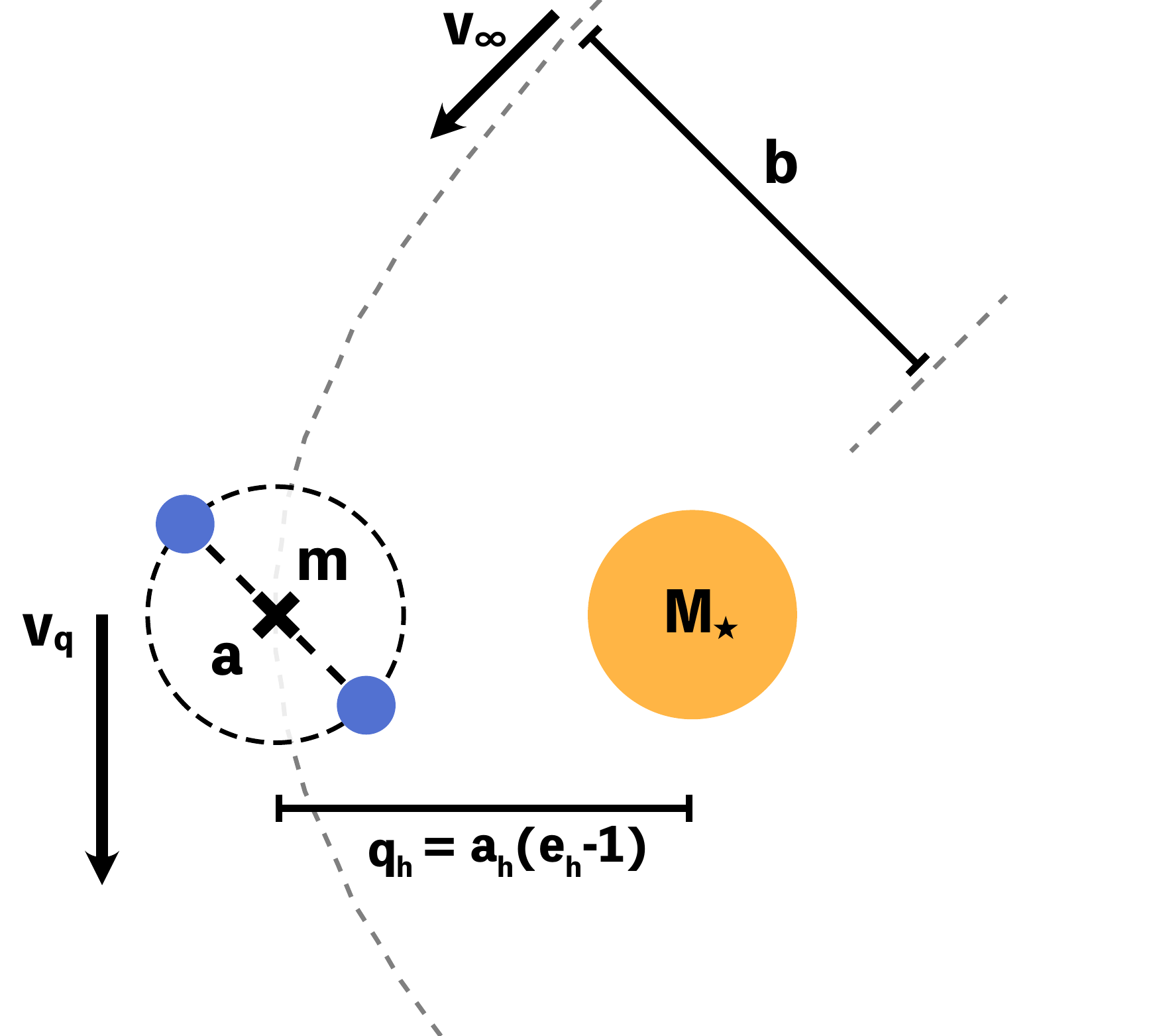}
  \caption{Hyperbolic trajectory of the binary planet relative to the passing star. Two planets (blue circles) with a total mass $m$ orbit their mutual barycenter (black cross) with the relative semimajor axis $a$. The periastron distance (closest approach) of the barycenter is denoted as $q_h$, which is related to the hyperbolic semimajor axis $a_h$ and eccentricity $e_h$. The geometry of the flyby trajectory is determined by the impact parameter $b$, the relative velocity between the binary barycenter and the star at infinity $v_\infty$, and the stellar mass $M_\star$.}
  \label{fig:jumbo-schematic}
\end{figure}

We consider a hyperbolic flyby of the binary barycenter relative to the passing star with a impact parameter larger than the binary semimajor axis ($b > a$). We do not consider $b < a$ flybys because they are much rarer. As illustrated in Figure~\ref{fig:jumbo-schematic}, the hyperbolic path can be described with the following orbital elements:
\begin{equation}\label{eq:a_q_e}
  \begin{aligned}
    a_h &= \frac{\mathcal{G}M_\star}{v_\infty^2}, \\
    q_h &= a_h(e_h - 1) = a_h\left(\sqrt{1 + j_h^2} - 1\right),\\
    j_h &= \frac{b}{a_h},
\end{aligned}
\end{equation}
where $\mathcal{G}$ is the gravitational constant. The hyperbolic semimajor axis $a_h$ and the periastron distance $q_h$ are determined by the impact parameter $b$, the relative velocity between the star and the binary barycenter at infinity $v_\infty$, and the mass of the passing star $M_\star$. Assuming $M_\star = 1 M_\odot$, the hyperbolic semimajor axis can be quickly estimated with $a_h = (30~\text{km}~\text{s}^{-1}/v_\infty)^2~\text{au}$. For young massive clusters with a typical velocity dispersion of $\sigma \sim 1$--$2~\text{km}~\text{s}^{-1}$, $a_h$ is of the order of $\sim$100--1000~au for solar-mass stars. The ratio between $b$ and $a_h$ is defined as $j_h$, representing the dimensionless angular momentum on a given semimajor axis \citep[see][]{Tremaine.2023dps}.

The pericenter velocity $v_q$ can be obtained via the conservation of angular momentum
\begin{equation}\label{eq:v_q}
    v_q = \frac{b v_\infty}{q_h}.
\end{equation}
The typical timescale of the encounter is \footnote{There are also other definitions for the encounter timescale, such as $b/v_\infty$ and $q_h/v_\infty$. However, we notice that these definitions differ significantly from the real interaction timescale, when the trajectory is near-parabolic ($e_h \sim 1$).}
\begin{equation} \label{eq:t_enc}
    t_\text{enc} \equiv \frac{2 q_h}{v_q} ,
\end{equation}
and a \textit{close} encounter is defined as $t_\text{enc} \lesssim P$, with the orbital period of the binary $P$ given by
\begin{equation}\label{eq:P}
    P = 2\pi\sqrt{a^3/(\mathcal{G}m)} .
\end{equation}
Here $a$ is the binary semimajor axis and $m$ is the binary total mass. During such encounters, the relative location of the two planets does not change over the span when the gravitational pull from the star is most intense. As a result, the change in the specific orbital energy of the binary, $E = -\mathcal{G}m/(2a)$, can be estimated with an impulse approximation. We do not consider tidal interaction beyond the impulse approximation, such as the quasi-resonant interaction \citep{DOnghia.2010}. 
The energy change induced by a passing perturber is given by \citep{Farinella.1993}:
\begin{equation}\label{eq:Delta-E-orig}
  \begin{aligned}
    \Delta E \simeq  \frac{\mathcal{G}^2 M_{\star} m P }{4 a  b v_\infty q_h } \int_{-f_l}^{f_l} \Psi \mathrm{d} f, 
\end{aligned}
\end{equation}
where $f$ is true anomaly and $f_l = \cos^{-1}({-1/e_h})$. Here, the integral term evaluates the total energy change along the hyperbolic path, which is related to the spatial orientation of the hyperbola relative to the binary orbital plane (ascending node $\Omega$ and inclination $i$) as well as the exact locations of the two components. Monte-Carlo simulations of random flyby angles and initial locations show that the orientation integral has an average of zero and a root mean square of the order of unity (see figure 3 of \citealt{Farinella.1993}). This implies that random close encounters have equal probabilities of softening and hardening the binary. Therefore, the effect of multiple flybys can be modeled as a random walk in $E$ space, with the typical energy change $|\Delta E|$ (i.e., the variance of the $\Delta E$ distribution) being
\begin{equation}\label{eq:Delta-E-magnitude}
  \begin{aligned}
    \left| \Delta E \right| &\simeq \frac{\mathcal{G}^2 M_{\star} m P }{4 a  b v_\infty q_h }.\\
\end{aligned}
\end{equation}

It is important to note that this energy random-walk approximation appears to conflict with Heggie's law, which states that soft binaries become progressively softer over time due to multiple stellar encounters \citep{Heggie.1975, Binney.2008}. However, we emphasize that binary planet -- star encounters are primarily influenced by tidal effects rather than close three-body interactions, rendering Heggie's law inapplicable in this context.

We define \textit{critical} flybys as stellar encounters with a typical energy change larger than the binary binding energy $\left| \Delta E \right| > \left|E\right|$. Such encounters would likely induce significant changes to the semimajor axis $a$ and orbital eccentricity $e$ of the binary, or even ionize it. Rearranging the expression, one obtains an inequality for two timescales:
\begin{equation}\label{eq:t_cr}
  \begin{aligned}
   t_\text{cr} \equiv \frac{2 q_h}{v_\infty} j_h <  P,
\end{aligned}
\end{equation}
where the left-hand side is defined as the critical flyby timescale $t_\text{cr}$. 

The encounter timescale $t_\text{enc}$ (Equation~\ref{eq:t_enc}) can be rewritten as $t_\text{enc}=(2q_h/v_\infty)(q_h/b)$ using Equation~\eqref{eq:v_q}. The relation between the two timescales, $t_\text{enc}$ and $t_\text{cr}$, is thus given by
\begin{equation}\label{eq:t_cr_t_enc}
  \begin{aligned}
    t_\text{cr} =\left(e_h + 1\right)  t_\text{enc} ,
\end{aligned}
\end{equation}
thus $t_\text{cr} > 2 t_\text{enc}$ holds for all hyperbolic paths. As a result, $t_\text{cr} < P$ is always a stronger constraint than $t_\text{enc} < P$; in other words, all critical flybys are close encounters where the impulse approximation is applicable.

\begin{figure}[htb!]
  \centering
  \includegraphics[width=1\columnwidth]{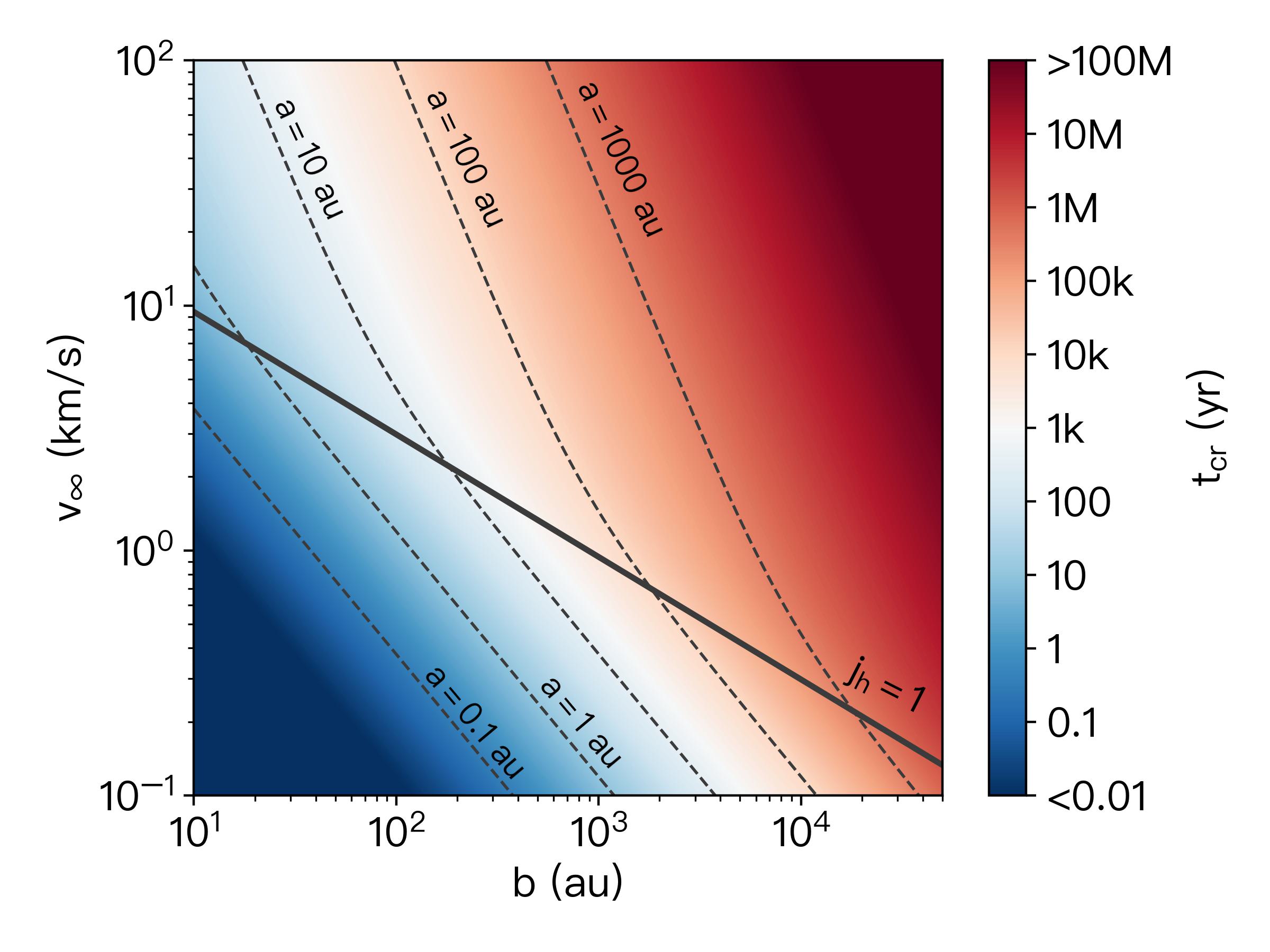}
  \caption{Critical timescale $t_\text{cr}$ (Equation~\ref{eq:t_cr}) computed on the $b$-$v_\infty$ space assuming a solar-mass star. Dashed lines denote the orbital periods of 10~$M_J$ JuMBOs with a wide range of semimajor axes $a$, of which critical flybys lie on the left-hand side. The solid line represents $j_h = 1$ ($e_h = \sqrt{2}$), which divides the parameter space into the hyperbolic regime ($j_h > 1$, above the line) and the near-parabolic regime ($j_h < 1$, below the line).}
  \label{fig:timescale}
\end{figure}

We compute the critical timescale for a series of $b$ and $v_\infty$ using Equation~\eqref{eq:t_cr} and plot the results in Figure~\ref{fig:timescale}. The critical flyby timescale spans several orders of magnitude, from a few days to more than 100~Myr. $t_\text{cr}$ also scales differently given the geometry of the flyby trajectory: Near-parabolic flybys with $e_h \sim 1$ ($j_h \sim 0$) affect the binary more effectively than hyperbolic flybys ($j_h > 1$) do, as indicated by a slope change separated by the solid line $j_h = 1$.

Next, we derive the scaling law relevant to the critical flyby criterion. One first approximates the pericenter distance
\begin{equation}\label{eq:q_h_expansion}
\begin{aligned}
   q_h & \approx
   \begin{cases}
     \cfrac{b^2}{2 a_h} \quad &\text{ if } j_h < 1, \\
     b  \quad &\text{ if } j_h > 1, \\
   \end{cases}
\end{aligned}
\end{equation}
and then explicitly writes the fractional orbital energy change for different $j_h$
\begin{equation}\label{eq:delta_Ecal}
| \Delta \mathcal{E} |  \equiv \frac{|\Delta E|}{|E|}
=\begin{cases}
     \left(\mathcal{G}M_\star \right)^2 P b^{-3} v_\infty^{-3}\quad & \text{ if } j_h < 1, \\
    \left(\mathcal{G}M_\star  P/2\right) b^{-2} v_\infty^{-1}\quad &\text{ if } j_h > 1.
  \end{cases}
\end{equation}
Equating $|\Delta \mathcal{E}|$ with 1 yields
\begin{equation}\label{eq:b_cr}
  b_\text{cr} =\begin{cases}
     \left[\left(\mathcal{G}M_\star \right)^2 P\right]^\frac{1}{3} v_\infty^{-1}\quad & \text{ if } j_h < 1, \\
   \left[\mathcal{G}M_\star  P/2 \right]^\frac{1}{2} v_\infty^{-\frac{1}{2}}\quad &\text{ if } j_h > 1,
  \end{cases}
\end{equation}
where $b_\text{cr}$ is the impact parameter for critical flybys. Notice $b_\text{cr} \propto v_\infty^{-1}$ in the near-parabolic regime and $b_\text{cr} \propto v_\infty^{-\frac{1}{2}}$ in the hyperbolic regime, which corresponds to the slope change in Figure~\ref{fig:timescale}.

One can easily estimate the orbital stability of a binary planet against a particular flyby by comparing $t_\text{cr}$ with $P$. In Figure~\ref{fig:timescale}, we plot orbital periods of 10~$M_J$ JuMBOs with $a$ spanning from 0.1~au to 1000~au in dashed lines, and the areas to the left of these lines indicate flyby parameters that would likely induce significant changes in binding energy. For JWST JuMBOs with $a = 25$--400~au, critical flybys occur at $b_\text{cr} \lesssim 600$--6,000~au for $v_\infty = 1$~$\text{km}$~$\text{s}^{-1}$, and $b_\text{cr} \lesssim 300$--3,000~au for $v_\infty = 2.5$~$\text{km}$~$\text{s}^{-1}$. It is worth mentioning that such close flybys may occur in some of the observed systems (see JuMBO33 and 34 in Figure 3 of \citealt{Pearson.2023}).

Another analytical criterion for the tidal disruption of a binary system can be derived using the Hill radius (also known as the tidal radius, \citealt{Hill.1878}):
\begin{equation}\label{eq:hill_radius}
\begin{aligned}
   R_\text{hill} \equiv q_h \left(\frac{m}{3M_\star}\right)^{\frac{1}{3}} < a,
\end{aligned}
\end{equation}
where $a$ is the semimajor axis of the binary. This criterion is widely employed in studies on the tidal disruption of binaries by supermassive black holes as a mechanism for creating hyper-velocity stars (see, e.g., \citealt{Hills.1988, Bromley.2006, Sari.2010}), as well as in the disruption of binary minor planets during Neptune scatterings \citep{Parker.2010}.

To compare our critical flyby criterion (Equation~\ref{eq:t_cr}) with the Hill criterion, we substitute the binary's semimajor axis $a$ in Equation~\eqref{eq:hill_radius} with its orbital period $P$, and reformulate the Hill criterion in terms of a timescale:
\begin{equation}\label{eq:hill_timescale}
\begin{aligned}
   t_\text{hill} \equiv \frac{\pi}{\sqrt{3}}  \left( e_h + 1 \right)^{\frac{1}{2}} t_\text{enc} < P.
\end{aligned}
\end{equation}

The ratio of the two timescales is then given by
\begin{equation}\label{eq:ratio}
\begin{aligned}
   \frac{t_\text{hill}}{t_\text{cr}} = \frac{\pi}{\sqrt{3}} \left( e_h + 1 \right)^{-1/2}.
\end{aligned}
\end{equation}
For near-parabolic flybys ($e_h \sim 1$), these two timescales are essentially equivalent. However, for hyperbolic, high-speed flybys ($e_h \gg 1$), the ratio scales as $t_\text{hill}/t_\text{cr} \propto e_h^{-1/2}$. Without resorting to numerical simulations, we argue that the critical timescale we derived is more accurate for predicting binary disruption in high-speed encounters compared to the Hill criterion. The Hill criterion (Equation~\ref{eq:hill_radius}) depends only on the periastron distance of the encounter, irrespective of the periastron velocity. This assumption is flawed because, in the high-speed limit, the encounter trajectory approximates a straight line, and the binary's energy change $\Delta E$ must be proportional to the ``interaction time'' of the tidal force along the path, leading to $\Delta E \propto v_\infty^{-1}$ (assuming $v_p \approx v_\infty$ for high-speed encounters). For a fixed periastron distance $q_h$, $v_\infty \propto \sqrt{e_h}$ at large $e_h$, so $\Delta E \propto e_h^{-1/2}$ must be expected from a realistic criterion. The Hill criterion, however, does not scale with eccentricity at high velocities, whereas $t_\text{cr} \propto e_h^{1/2}$ (Equation~\ref{eq:t_cr_t_enc}). Thus, this simple analysis demonstrates that our critical flyby criterion represents an improvement over the more commonly used Hill criterion (or tidal radius), particularly for high-speed encounters at large eccentricities.

\subsection{Dynamical Lifetime Inside a Cluster}

Until now, we have only looked at the average effects caused by a single passing star. However, in a cluster environment where thousands of stars are present, it is essential to consider the ongoing interactions from multiple stars, each with varying trajectories and characteristics.

The frequency of encounters with the parameters $b$ and $v_\infty$ for a period of time $t$ is given by 
\begin{equation}\label{eq:enc_freq}
N = n_\star t (\pi b^2) v_\infty,
\end{equation}
where $n_\star$ is the number density of the star cluster, and $\pi b^2$ represents the encounter cross section. For critical flybys with $b < b_\text{cr}$, the induced fractional energy change $\Delta \mathcal{E}$ is of order unity. In contrast, for non-critical distant flybys, the energy drift can accumulate, with the cumulative change after $N$ flybys following the random walk equation $|\Delta \mathcal{E}|_{N} = \sqrt{N} |\Delta \mathcal{E}|$. Notably, $|\Delta \mathcal{E}|_{N}$ scales as $b^{-2}$ in the near-parabolic case and $b^{-1}$ in the hyperbolic case. This implies that in both scenarios, the total energy change is dominated by the single closest stellar encounter rather than by the cumulative effects of more frequent, distant encounters. This contrasts with the case of soft binary stars, where cumulative energy kicks (evaporation) rather than a single close encounter (ejection) are the primary mechanism for binary disruption \citep{Binney.2008}. This difference arises from the distinct assumptions inherent in each scenario. As a result, the ionization of a planetary-mass binary mostly results from the closest encounter, and the dynamical lifetime can be defined as the expected time interval between two critical flybys (Equation~\ref{eq:T_life}).

\begin{figure*}[htb!]
\begin{equation}\label{eq:T_life}
\begin{aligned}
T &\equiv \frac{1}{n_\star (\pi b_\text{cr}^2) v_\infty} \\
&\simeq  \begin{cases} 0.9~\text{Myr}~  \left(\cfrac{n_\star}{10^4~\text{pc}^{-3}}\right)^{-1} \left(\cfrac{M_\star}{M_\odot}\right)^{-\frac{4}{3}} \left(\cfrac{m}{10~M_J}\right)^{\frac{1}{3}}  \left(\cfrac{a}{100~ \text{au}}\right)^{-1} \left(\cfrac{v_\infty}{1~\text{km s}^{-1}} \right) \quad & \text{ if } j_h < 1, \\

1.4~\text{Myr}~\left(\cfrac{n_\star}{10^4~\text{pc}^{-3}}\right)^{-1} \left(\cfrac{M_\star}{M_\odot}\right)^{-1} \left(\cfrac{m}{10~M_J}\right)^{\frac{1}{2}}  \left(\cfrac{a}{100~ \text{au}}\right)^{-\frac{3}{2}}\quad & \text{ if } j_h > 1.
\end{cases}
\end{aligned}
\end{equation}
\end{figure*}

Notice that the discontinuity between the two equations at $j_h = 1$ arises from different approximations of $q_h$ in Equation~\eqref{eq:q_h_expansion}. Adopting the typical number density of $1\times10^4~\text{pc}^{-3}$ and the velocity dispersion of $1~\text{km s}^{-1}$ for young star clusters, Equation~\eqref{eq:T_life} suggests that the dynamical lifetime of a JuMBO with $a = 100$~au is only on the order of $\sim$1~Myr, consistent with the recent full N-body numerical simulations \citep{Portegies-Zwart.2023}. It should be noted that the structure of star clusters dynamically evolves, so the JuMBO stability may vary depending on the cluster density over time.

We also calculate the dynamical lifetime of potential JuMBOs within other types of cluster, assuming a constant star density and relative velocity. For open clusters such as Hyades ($n_\star = 1~\text{pc}^{-3}$ and $v_\infty = 0.25~\text{km s}^{-1}$, \citealt{Perryman.1997}), binary planets similar to the JWST JuMBOs are stable for several billion years, significantly longer than the $\sim$700~Myr age of the Hyades; Conversely, the same type of binary planets would have long been destabilized if they formed in the globular cluster 47 Tucanae ($n_\star = 1\times10^5~\text{pc}^{-3}$ and $v_\infty = 10~\text{km s}^{-1}$, \citealt{Meylan.1986}), given their short dynamical lifetime ($\lesssim$1~Myr) compared to the cluster age of 11~Gyr. Interestingly, if any JuMBOs escape stellar clusters and reside in the galactic disk, their dynamical lifetimes are generally much longer than the age of the universe, given a stellar number density of $n_\star \approx 0.01~\text{pc}^{-3}$ and a typical encounter velocity of $v_\infty = 30~\text{km s}^{-1}$.

Finally, we present the tidal radius of a binary planet within star clusters, assuming a Plummer model potential \citep{Binney.2008}:
\begin{equation}\label{eq:tidal_radius}
    R_\text{tide} = \left(\frac{m}{3 M_\text{tot}}\right)^{1/3} \left(r^2 + b_\text{plu}^2\right)^{1/2},
\end{equation}
where $M_\text{tot}$ is the total mass of the cluster, $b_\text{plu}$ is the Plummer scale length, and $r$ is the radial distance from the cluster center. The tidal radius defines the maximum separation at which a binary planet is disrupted by the cluster's gravitational potential. By adopting $M_\text{tot} = 900 M_\odot$ and $b_\text{plu} = 0.5~\text{pc}$ for the Trapezium Cluster \citep{Portegies-Zwart.2023}, and assuming $r =0$--0.6~pc, we find tidal radii of 700--1200~au for $1 M_J$ and 1600--2500~au for $10 M_J$ JuMBOs, respectively.

\subsection{Numerical Validation}

We carry out numerical simulations in which binary planets encounter a solar-mass star from random directions. For each simulation, one thousand equal-mass binaries with a binary mass of $10 M_J$ are initialized on a circular orbit with $a_0 = 15$~au. Their barycenter is then set on a hyperbolic trajectory relative to the passing star with $v_\infty = 1~\text{km}~\text{s}^{-1}$. Both the initial binary mean anomalies and the longitude of the ascending nodes are distributed uniformly in $(0,2\pi)$, while the relative inclinations between the hyperbolic trajectory and the binary orbital plane are distributed uniformly in $\cos(i)$, corresponding to the isotropic angles of entry. We then obtain the corresponding critical impact parameter $b_\text{cr} = 460$~au using Equation~\eqref{eq:b_cr}, and set $b = b_\text{cr}/1.5, b_\text{cr},$ and $1.5b_\text{cr}$, respectively, for the three experiments. All three cases have $j_h < 1$ and thus belong to the near-parabolic regime. Simulations are performed using the \textsc{ias15} integrator of \textsc{rebound} \citep{Rein.2014}.

\begin{figure*}[htb!]
  \centering
  \includegraphics[width=1.0\textwidth]{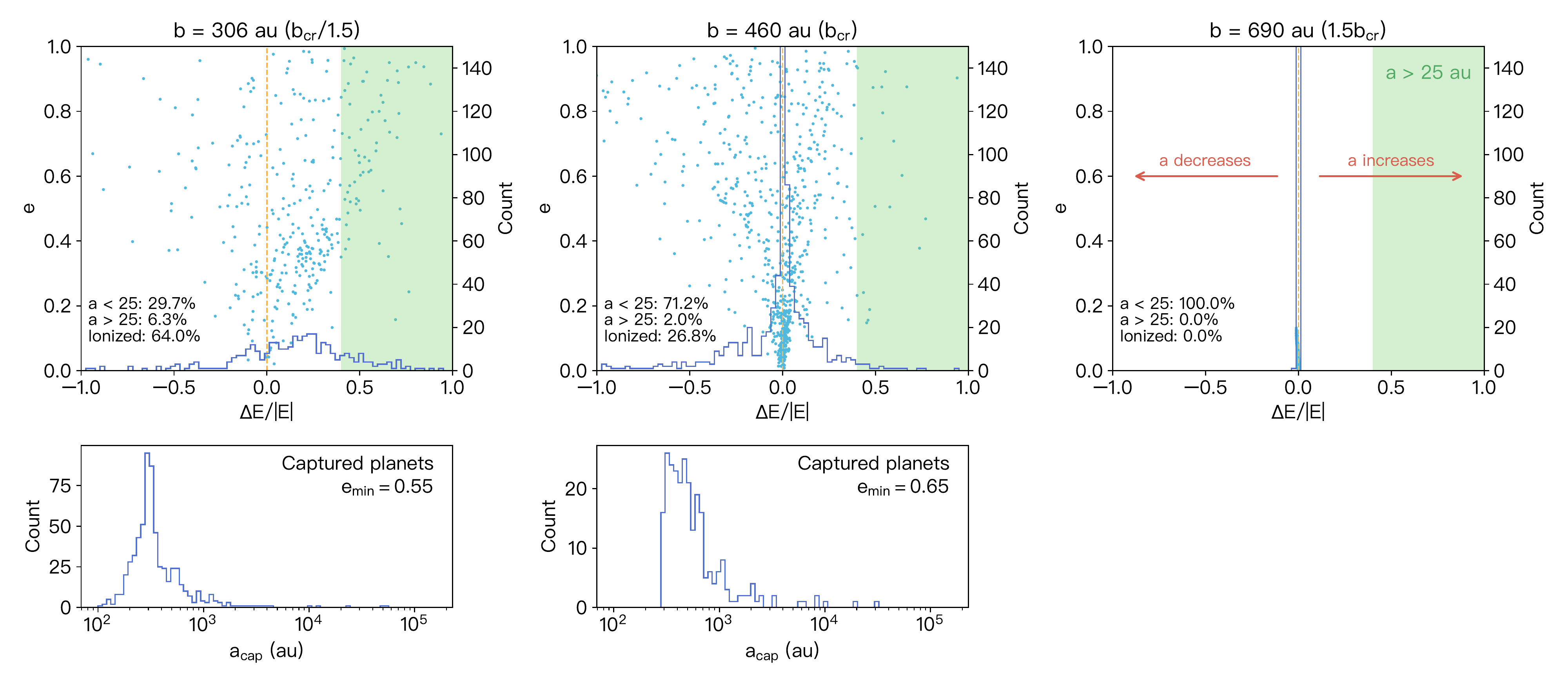}
  \caption{\textbf{Upper:} Orbital distributions and histograms of 1,000 equal-mass binary planets after encountering a solar-mass star with $v_\infty = 1~\text{km}~\text{s}^{-1}$. Binaries are initialized with $a_0 = 15$~au, $e_0 = 0$, and $m = 10~M_J$, while the chosen impact parameters correspond to a deep critical flyby ($b_\text{cr}/1.5$, left panel), a critical flyby ($b_\text{cr}$, middle panel), and a distant flyby ($1.5b_\text{cr}$, right panel), respectively. The $x$ axis indicates the fractional energy change $\Delta \mathcal{E}$. The left $y$ axis represents the binary eccentricity post flyby, whereas the right $y$ axis denotes the number of binaries in each bin. The orange dashed line marks the energy change of 0, with a negative $\Delta \mathcal{E}$ corresponding to a decrease in $a$, and vice versa (red arrows). The green shaded region denotes the observable JuMBOs with $a > 25$~au. \textbf{Lower:} Histograms of semimajor axes of captured planets ($a_\text{cap}$) from ionized JuMBO pairs. }

  \label{fig:simulations}
\end{figure*}

The post-flyby orbital distributions of the three simulations are shown in Figure~\ref{fig:simulations}, with $\Delta \mathcal{E}$ denoting the fractional energy change and $e$ denoting the eccentricity. We decide to plot $\Delta \mathcal{E}$ instead of $\Delta a$ because it is a dimensionless scale-free quantity. In other words, even if we ran the same simulations (with corresponding $b_\text{cr}$) for binaries with different $m$ and $a_0$, their orbital distributions in the $\Delta \mathcal{E}$ space would still look similar to those in Figure~\ref{fig:simulations}.

We define the following branching fractions:
\begin{enumerate}
    \item $f_{<\text{25}}$ --- the number of binaries with $a < 25$~au divided by the total number of primordial binaries,
    \item $f_{>\text{25}}$ --- the number of binaries with $a > 25$~au divided by the total number, and
    \item $f_\text{ion}$ --- the number of ionized binaries (signified by a positive binary energy post flyby) divided by the total number (i.e., \textit{ionization} fraction).
\end{enumerate}
With the above definitions, we have $f_{<\text{25}} + f_{>\text{25}} + f_\text{ion} = 1$. The values of these three fractions are shown in Figure~\ref{fig:simulations}.

The left and middle panels of Figure~\ref{fig:simulations} demonstrate the aftermath of critical flybys with different impact parameters. Although both flybys show roughly equal chances of softening and hardening the orbit (that is, a near symmetric distribution centered at $0$), the middle panel shows a strong peak at $\Delta \mathcal{E} = 0$, implying that a significant portion of binaries are not strongly affected by the flyby. In comparison, 64\% of the binaries are ionized by deeper critical flybys.

When the equal-mass binaries are ionized by the passing star, a common outcome is the capture of one planet by the passing star and the ejection of the other planet\footnote{Additional simulations with unequal-mass binaries were also carried out, and the result suggests a lower capture probability.}. The captured planets are on highly eccentric orbits around the star, with $a_\text{cap}$ peaked at 300--400~au and the minimal eccentricity of $e_\text{min} = 0.5$--0.6 (lower panels in Figure~\ref{fig:simulations}). These distant and elliptic planetary orbits are mostly not stable inside a dense cluster environment, and one can expect that subsequent stellar perturbations would easily disrupt them and produce FFPs \citep{Spurzem.2009}. It is worth noting that three-body interactions occurred frequently in the early Solar System, both among planetesimals \citep{Funato.2004} and between planetesimals and a planet. For example, Triton -- the largest moon of Neptune -- is thought to be captured during a close encounter between Neptune and a transneptunian binary \citep{Agnor.2006}.

There is an apparent dichotomy between critical flybys and noncritical flybys. As illustrated in the right panel of Figure~\ref{fig:simulations}, the noncritical ($b = 1.5b_\text{cr}$) flybys induce little variation in both the binary energy and eccentricity. In a cluster environment, such flybys would be roughly twice as likely as critical ($b_\text{cr}$) flybys (Equation~\ref{eq:enc_freq}). However, the cumulative effect of two $1.5b_\text{cr}$ flybys is by no means comparable to a $b_\text{cr}$ flyby. This again validates our previous analysis that the orbital change of a binary planet is dominated by the single closest passing star.

\section{Application to JWST JuMBOs}\label{sec:ratio}

We have demonstrated that starting from a single population, gravitational interactions with passing stars are able to produce a variety of binary systems from a few to several hundred au. The same dynamical process also produces planetary binaries on tighter orbits (close binaries), FFPs, and temporarily captured planets, which are expected to be liberated as FFPs shortly after due to their extremely large semimajor axes.

Limited by the observational capability of the telescope, however, only wide binaries with separations above a certain threshold ($25$~au, for JuMBOs in the Trapezium) can be resolved by JWST, and it is unclear whether the `non-binary' PMOs are isolated FFPs, or unresolved close JuMBOs, or more likely, a combination of both. The applicable constraint from \citet{Pearson.2023} is the $\sim$9\% JuMBO-PMO fraction (the number of $a > 25$~au JuMBOs divided by the total number of all planetary-mass objects), i.e., the wide binary fraction $f_\text{wide}$, which is also given by
\begin{equation}\label{eq:wide_binary_fraction}
\begin{aligned}
f_\text{wide} &= \frac{f_{>25}}{2 \times f_\text{ion} + f_{<25} + f_{>25} + R_{S:B}} \\
&=\frac{f_{>25}}{1 + f_\text{ion}+R_{S:B}},
\end{aligned}
\end{equation}
where $R_{S:B} = N_\text{single}/N_\text{binary}$ is the primordial ratio between single planets and binary planets governed by the formation process. Since the formation mechanism is still unknown, we simply assume that it only produces binary objects, that is, $R_{S:B} = 0$. As a result, Equation~\ref{eq:wide_binary_fraction} gives the upper bound of $f_\text{wide}$.

The ionization fractions for the two critical flybys presented in Figure~\ref{fig:simulations} are 64\% and 27\%, respectively. Deeper critical flybys are also more efficient in making $a > 25$~au binaries (6.3\% compared to 2.0\%), because of stronger energy kicks. The resultant $f_\text{wide}$ therefore ranges from 2\% to 4\%. It is worth pointing out that $f_\text{wide}$ is very sensitive to the initial $a_0$ and the definition of wide JuMBOs. Therefore, we next explore numerically what initial $a_0$ best reproduces the $\sim$9\% wide binary fraction, assuming a ``primordial JuMBO population'' formed at the beginning of the star cluster.

Instead of considering only the single closest encounters, we adopt a Monte-Carlo framework from \citet{Heisler.1986} to numerically simulate the evolution of JuMBOs undergoing multiple encounters, in which hyperbolic trajectories are initialized using a realistic impact parameter distribution ($\mathrm{d}N/\mathrm{d}b \propto b$, with $b_\text{max} = 10,000$~au) and a Maxwellian distribution for $v_\infty$, with the most probable encounter velocity being $2~\text{km s}^{-1}$ \citep{McCaughrean.1994}. For each encounter, we calculate the corresponding $b_\text{cr}$ using Equation~\ref{eq:b_cr} and integrate only those flybys where $b/b_\text{cr} < 2$ is satisfied\footnote{Test runs that included all distant encounters showed that flybys with $b/b_\text{cr} > 2$ induce fractional energy changes on the order of $\sim 10^{-10}$ at these encounter velocities. Therefore, excluding such encounters does not significantly affect the results.}. 

\begin{figure*}[htb!]
  \centering
  \includegraphics[width=1.0\textwidth]{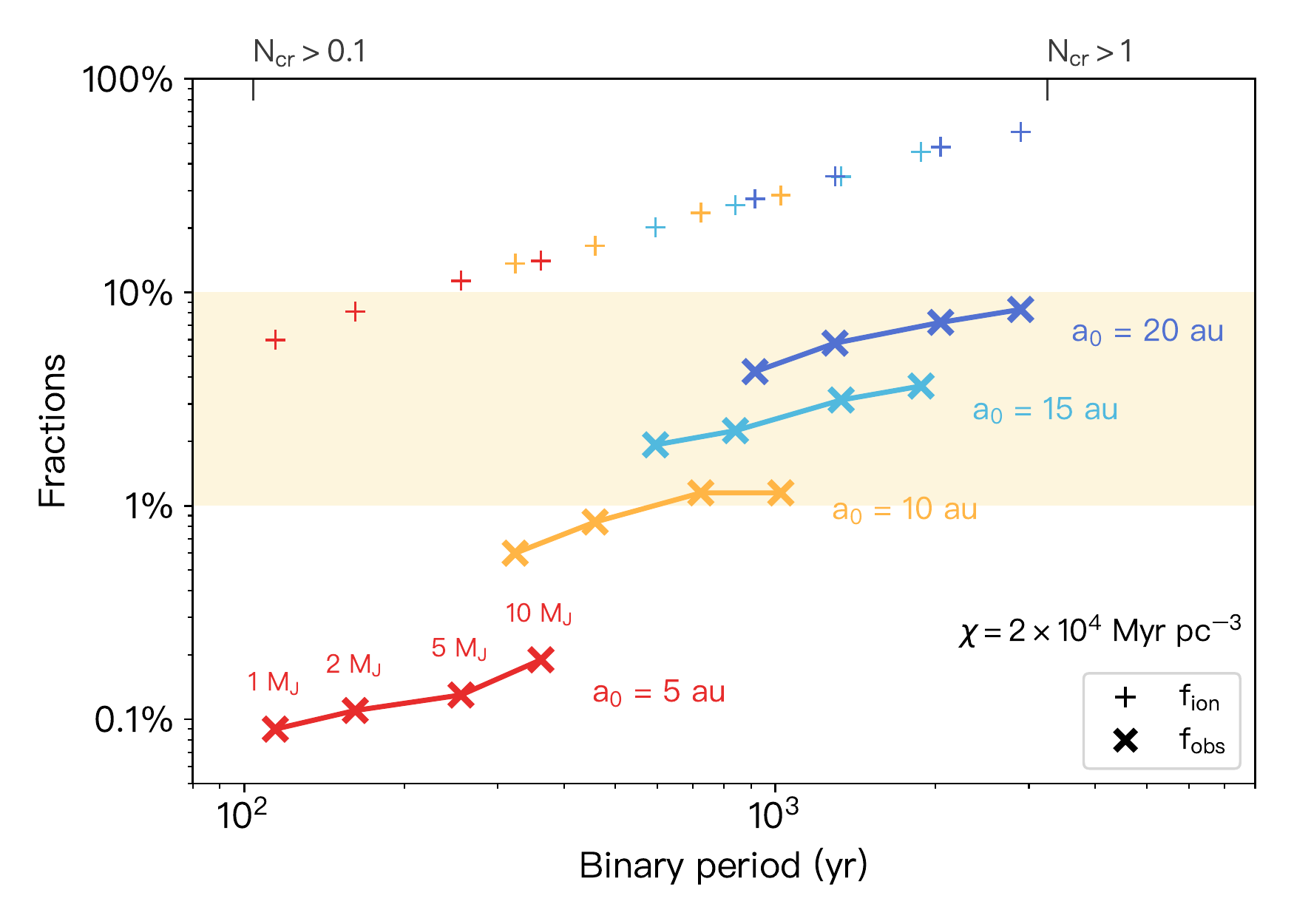}
  \caption{\textbf{Left:} Ionization fraction $f_\text{ion}$ ($+$) and wide binary fraction $f_\text{wide}$ ($\times$) from stellar flyby simulations as a function of the binary period $P$. Each symbol represents the result of 5,000 JuMBOs with the same mass (from right to left, $m=1, 2, 5$, and 10~$M_J$, respectively) and $a_0$ (grouped by the same color) interacting with random solar mass stars multiple times. The impact parameter $b$ is generated from a realistic distribution, assuming a density-weighted residence time of $\chi = 2 \times 10^4$~Myr~pc$^{-3}$, and $v_\infty$ follows a Maxwellian distribution. The range of $f_\text{wide}$ potentially consistent with the JWST discovery (1\%--10\%) is highlighted in yellow. \textbf{Right:} Histograms of semimajor axes and eccentricities of wide JuMBOs ($a > 25$~au) generated from multiple stellar flybys. The red dashed line indicated a power law scaling of $\mathrm{d}N/\mathrm{d}a \propto a^{-2.5}$.}
  \label{fig:f_obs_all}
\end{figure*}

After each encounter, the semi-major axis $a$ and eccentricity $e$ of the binary are recorded and used as the initial conditions for the subsequent flyby. We define a single parameter $\chi = \int_0^{t} n_\star \mathrm{d}t$ as the stellar number density-weighted residence time \citep{Batygin.2020jr}, which indicates the time-accumulated effect taking into account the evolution of the cluster. We do not consider other factors, such as the evolution of velocity dispersion, in the context of an evolving cluster, as addressing these would necessitate conducting a full N-body simulation. We set $\chi = 2\times 10^4$~Myr~pc$^{-3}$, which corresponds to the lower bound of the Trapezium Cluster core density \citep{Heggie.1992, Hillenbrand.1998}.  This cluster environment results in a mean encounter frequency of $\approx$2~kyr$^{-1}$ for $b_\text{max} = 10,000$~au flybys (see Equation 4 of \citealt{Heisler.1986}), and the mean number of $b/b_\text{cr} < 2$ flybys (which we actually integrate) per binary ranges from 0.5 to 5, depending on the initial $a_0$ and mass of the binary.

The results of the multiple-encounter simulation are presented in Figure~\ref{fig:f_obs_all}. In the left panel, pluses ($+$) and crosses ($\times$) represent $f_\text{ion}$ and $f_\text{wide}$ from a simulation set in which 5,000 binary planets randomly encounter passing stars multiple times. The $x$ axis denotes the binary period $P$, a proxy for its dynamical stability (see Equation~\ref{eq:t_cr}). The ionization fraction $f_\text{ion}$ grows monotonically as the binary period, in agreement with analytical theory.

For primordially tight JuMBOs ($a_0 < 25$~au), orbital expansion is the main cause for generating wide binaries ranging from 25 to 400~au. The post-evolution $f_\text{wide}$ lies between 1\% to 10\% (yellow highlight) for $a_0 = 15$ and 20~au, consistent with the $\sim$9\% observed by JWST in terms of order of magnitude. $f_\text{wide}$ is merely $\sim$0.1\% for $a_0 = 5$~au JuMBOs, due to stronger orbital stability (signified by their low ionization fractions) and fewer chances of getting wide orbits. These two effects can be quantitatively estimated in the following manner: 1) To increase a binary $a$ from 5~au to $>$25~au, a fractional energy change of $\Delta \mathcal{E} = (0.8,1)$ is required. Compared to $\Delta \mathcal{E} = (0.4,1)$ needed for raising 15~au to $>$25~au, it is an event at least a factor of three (0.6/0.2) less likely to occur with random energy kicks. 2) Critical flybys are roughly $5\times$ more frequent for $a_0 = 15$~au than for $5$~au (obtained by combining Equations~\ref{eq:b_cr} and \ref{eq:enc_freq}). Therefore, we expect a factor of $\sim$15 difference in $f_\text{wide}$ for $a_0 = 15$~au and $5$~au, in excellent agreement with Figure~\ref{fig:f_obs_all}. There is also a factor of $\sim$2 difference in $f_\text{wide}$ for JuMBOs with different masses: more massive JuMBOs tend to have a smaller wide binary fraction, in line with JWST observation \citep[Figure 4 in][]{Pearson.2023}. In contrast, Equation~\eqref{eq:T_life} shows that if JuMBOs primordially formed as wide binaries ($a_0 > 25$~au), the more massive JuMBOs (which are more stable against stellar flybys) would tend to have higher $f_\text{wide}$, contradicting the JWST observation.

Additionally, we present the histograms of the semimajor axes and eccentricities of wide JuMBOs generated from multiple stellar encounters in the right panel of Figure~\ref{fig:f_obs_all}, color-coded by their initial semimajor axes $a_0$. The semimajor axis distribution of the resulting wide binary planets closely follows a power law of $\mathrm{d}N/\mathrm{d}a \propto a^{-2.5}$, with a median of $a \sim 30$~au. This is expected, as the production process resembles an energy random walk, making large-$a$ binaries less likely to form due to random energy kicks. A similar power-law distribution in semimajor axis (established by random energy kicks in planetary scatterings) is observed in the Kuiper Belt \citep{Beaudoin.2023} and in simulations related to the formation of the Oort Cloud \citep{Duncan.1987, Dones.2004}. Notably, \citet{Pearson.2023} reported a flat distribution of separations for 42 wide JuMBOs. A simple Kolmogorov-Smirnov test indicates that the $a^{-2.5}$ distribution is rejected with more than 99.9\% confidence when fitted to the full sample. However, if the sample is restricted to $a < 250$~au, containing 62\% (26/42) of the binary objects, the $a^{-2.5}$ distribution is not rejected ($p = 11\%$). This suggests that the majority of JuMBOs with $a > 250$~au either do not have bound orbits and merely appear close to each other, or they formed through a different mechanism than the stellar flyby scenario considered in this work. Future observations that characterize their proper motions and mutual orbits are warranted.

JuMBOs formed through stellar flybys typically exhibit a highly excited eccentricity distribution, with a median of $e \sim 0.5$, though they do not reach a thermal distribution ($\mathrm{d}N/\mathrm{d}e \propto 2e$). While eccentricity evolution plays a significant role in the formation of low-mass X-ray binaries (see, e.g., \citealt{Michaely.2016}), wide JuMBOs formed via stellar encounters generally do not achieve the extreme eccentricities where mutual tides and collisions become significant factors. This is again due to the fact that the evolution of a binary planet is dominated by the single closest flyby star.

Therefore, we conclude that if a primordial JuMBO population formed within the Trapezium Cluster, the scenario in which they initially resided on tight orbits ($a_0 \sim 10$--20~au) best reproduces the observed ratio. This scenario necessitates a high density-weighted residence time of $\chi \gtrsim 10^4$~Myr~pc$^{-3}$, implying that primordial JuMBOs likely spent most of their time near the cluster's core, where critical flybys significantly altered their orbital properties. The distributions of both $a$ and $e$ produced by our models offer a potential observational test of this formation scenario. This hypothesized population may also connect to tighter JuMBOs ($a < 5$~au; see \citealt{Beichman.2013} and \citealt{Best.2017}), and our results suggest that the current population of PMOs observed by JWST may include a portion of unresolved binaries. Follow-up observations that characterize the properties of these PMOs will be key to uncovering their origins.

\section{Discussion}\label{sec:dis}

In this work, we have built analytical tools to study the orbital evolution and dynamical stability of binary planets within star clusters. By conducting three-body simulations of random encounters between binary planets and stars, we have shown that critical stellar flybys are efficient at producing binary planets with a wide range of $a$ and $e$ from a dominant population. 

Applying our results to recently discovered JWST JuMBOs \citep{McCaughrean.2023}, we show that their typical dynamical lifetime is comparable to the age of the Trapezium cluster, suggesting that these wide binaries may have encountered stellar flybys and thus dynamically evolved. We propose that the JWST JuMBOs were formed as tighter binaries near/within the core of the Trapezium cluster. To best reproduce the observed $\sim$9\% wide binary fraction, an initial semimajor axis of $a_0 \sim 10$--20~au and a density-weighted residence time of $\chi \gtrsim 10^4$~Myr~pc$^{-3}$ are favored. 

The discovered JWST JuMBOs are \emph{currently} within a radius of $\sim$0.6~pc, larger than the core of the Trapezium cluster, which has a radius of $\sim$0.1--0.2~pc and a central density of $\sim$2--5$\times10^4$~pc$^{-3}$ \citep{McCaughrean.1994, Hillenbrand.1998}. Given that the dynamical timescale of the ONC (as well as the inner Trapezium Cluster) is comparable to its age, it is likely that the dynamical equilibrium of the cluster was either recently reached or not yet established \citep{Portegies-Zwart.2010}. Furthermore, three-body simulations of young massive clusters suggest that the stars were more clustered at an early stage, leading to a higher stellar density \citep[see Figure 1 of][]{Fujii.2015}. Taking into account the dynamical evolution of the cluster itself, it is thus reasonable to postulate that the discovered JWST JuMBOs have experienced $\chi \gtrsim 10^4$~Myr~pc$^{-3}$ in the past $\sim$1~Myr. 

The current JWST JuMBO distribution beyond the cluster core could also be produced by binary--star encounters: JuMBOs that had critical stellar encounters in the past are also those that had experienced a significant barycentric velocity kick. Specifically, the 90$^\circ$ deflection impact parameter $b_\text{90}$ has exactly the same definition as $a_h$ \citep{Binney.2008}, suggesting that any stellar encounters in the near-parabolic regime ($b < a_h$) would also deflect the relative velocity vector by $>$90$^\circ$. This would induce a barycentric velocity change comparable to its original speed. Therefore, even in a scenario where JuMBOs are hypothesized to have formed closer to the core, it is not so surprising to discover escaped members within 1 pc around the core at 1~Myr, given the typical barycentric velocity of $v = 1~\text{km s}^{-1} \approx 1~\text{pc Myr}^{-1}$. Future optical surveys on the outskirts of the Trapezium Cluster should help test this validity of this scenario.

During the encounters between binary planets and the flyby stars, one of the planets in the binary may get captured by the star into a wide-separation and eccentric orbit. Our simulations show that the production rate of such captured planets is not negligible, especially in the dense environment. This can potentially provide an alternative explanation for the existence of Jupiter-mass planets on extremely wide and eccentric orbits that have been found by direct imaging surveys \citep[e.g.,][]{Kraus.2014, Bryan.2016}. It may also be connected to the temporarily present rogue planet \citep{Gladman.2006, Huang.2022ml}, or the hypothesized ``Planet Nine'' \citep{Batygin.2019}, in the solar system.

\vspace*{3mm}

\section*{Acknowledgments}

We thank Subo Dong, Zenghua Zhang, Xiaochen Zheng, Stephen Kane, and Wei Xing for useful discussions. We thank the anonymous referee for a thorough and insightful report that led to an improved manuscript. Y.H. acknowledges funding support from Tsinghua University, NAOJ, and a grant from the Hayakawa Satio Fund awarded by the Astronomical Society of Japan. This work is supported by the National Science Foundation of China (Grant No. 12173021 and 12133005), the CASSACA grant CCJRF2105.
E. K. is supported by JSPS KAKENHI Grant Number 18H05438. We also acknowledge the Tsinghua Astrophysics High-Performance Computing platform for providing computational and data storage resources.

\textit{Software:} Rebound \citep{Rein.2012}, Numpy \citep{Harris.2020}, Matplotlib \citep{Hunter.2007}

\bibliography{ref_fixed}{}
\bibliographystyle{aasjournal}

\end{document}